\documentclass[runningheads]{llncs}
\usepackage[T1]{fontenc}
\usepackage{graphicx}

\usepackage{array}            % for better array/tabular formatting
\usepackage{multirow}         % for multirow cells in tables
\usepackage{pifont}           % for \ding{51} and \ding{55} symbols (checkmark and X)
\usepackage{caption}          % for adjusting the caption font size
\usepackage{float}

\begin{document}
\title{DINeuro: Distilling Knowledge from 2D Natural Images via Deformable Tubular Transferring Strategy for 3D Neuron Reconstruction}
\titlerunning{DINeuro: Distilling 2D Natural Knowledge for 3D Neuron Reconstruction}
\author{Yik San Cheng\inst{1} \and
Runkai Zhao\inst{1} \and
Heng Wang\inst{1} \and
Hanchuan Peng\inst{2} \and
Yui Lo\inst{1, 3, 4} \and
Yuqian Chen\inst{3, 4} \and
Lauren J. O’Donnell\inst{3, 4} \and
Weidong Cai\inst{1}}
\authorrunning{Y. S. Cheng, R. K. Zhao, H. Wang et al.}
\institute{School of Computer Science, The University of Sydney, Sydney, Australia \and
SEU-ALLEN Joint Center, Institute for Brain and Intelligence, Southeast University, Nanjing, China \and
Harvard Medical School, Boston, MA, USA \and
Brigham and Women’s Hospital, Boston, USA}

% \url{http://www.springer.com/gp/computer-science/lncs}
%
\maketitle              % typeset the header of the contribution
\begin{abstract}
Reconstructing neuron morphology from 3D light microscope imaging data is critical to aid neuroscientists in analyzing brain networks and neuroanatomy. With the boost from deep learning techniques, a variety of learning-based segmentation models have been developed to enhance the signal-to-noise ratio of raw neuron images as a pre-processing step in the reconstruction workflow. However, most existing models directly encode the latent representative features of volumetric neuron data but neglect their intrinsic morphological knowledge. To address this limitation, we design a novel framework, namely \textbf{\textit{DINeuro}}, that distills the prior knowledge from DINO which pre-trained on extensive 2D natural images to facilitate neuronal morphological learning of our 3D Vision Transformer. To bridge the knowledge gap between the 2D natural image and 3D microscopic morphologic domains, we propose a deformable tubular transferring strategy that adapts the pre-trained 2D natural knowledge to the inherent tubular characteristics of neuronal structure in the latent embedding space. The experimental results on the Janelia dataset of the BigNeuron project demonstrate that our method achieves a segmentation performance improvement of 4.53\% in mean Dice and 3.56\% in mean 95\% Hausdorff distance.

% 3D Neuron Reconstruction, Volumetric Image Segmentation, Transfer Learning, Vision Transformer, 3D Microscope Image 

\keywords{3D Neuron Reconstruction \and Volumetric Image Segmentation \and Transfer Learning \and Vision Transformer \and 3D Microscope Image.}
\end{abstract}
\section{Introduction}
 Reconstructing the 3D structure of single neurons plays a fundamental role in analyzing and understanding brain functionality and mechanisms~\cite{manubens2023bigneuron,liu2016rivulet,peng2021morphological,liu2024single,gao2023single,qiu2024whole}. This intricate process traces the tree-like neuron morphology and digitizes the neuronal structure from light microscopic images. Nevertheless, this task has consistently remained challenging. Historically, neuron reconstruction heavily relied on manual annotation by neuroscience specialists, which requires significant time and effort~\cite{wang2021ai,liu2018automated}. Directly reconstructing neuronal morphology through various conventional tracing techniques often yields suboptimal outcomes~\cite{wang2018memory} due to the structural complexity of neurons, as well as the inherent noises and discontinuities present in light microscope imaging data. 

The emergence of deep learning techniques has led to a significant rise in learning-based methods to automatically generate precise neuron segmentation images. This segmentation process serves as a pre-processing step in neuron reconstruction, forming a higher-quality foundation for tracing methods to achieve accurate reconstruction results. Recent advancements of these methods focus on accurately segmenting neuron foreground pixels through data-driven approaches. Notable innovations include multi-scale kernel fusion~\cite{Wang_2019_CVPR_Workshops}, homogeneous model knowledge transfer~\cite{wang2019segmenting}, and atrous spatial pyramid pooling to improve spatial resolution~\cite{li20193d}. Additional approaches leverage intrinsic semantic learning across volumes~\cite{wang2021voxel}, global graph reasoning~\cite{wang2021single}, and 3D point geometry learning~\cite{zhao2023pointneuron}. Despite these advances, the scarcity of high-quality neuron datasets and SWC annotations limits performance. The study in~\cite{cheng2024boosting} introduced a tailored weight transferring strategy to address this issue. However, most existing models focus on directly encoding latent representative features of volumetric neuron images but neglect to extract the intrinsic structural characteristics of neurons and integrate them to enhance segmentation models.

\begin{figure}
\includegraphics[width=\textwidth]{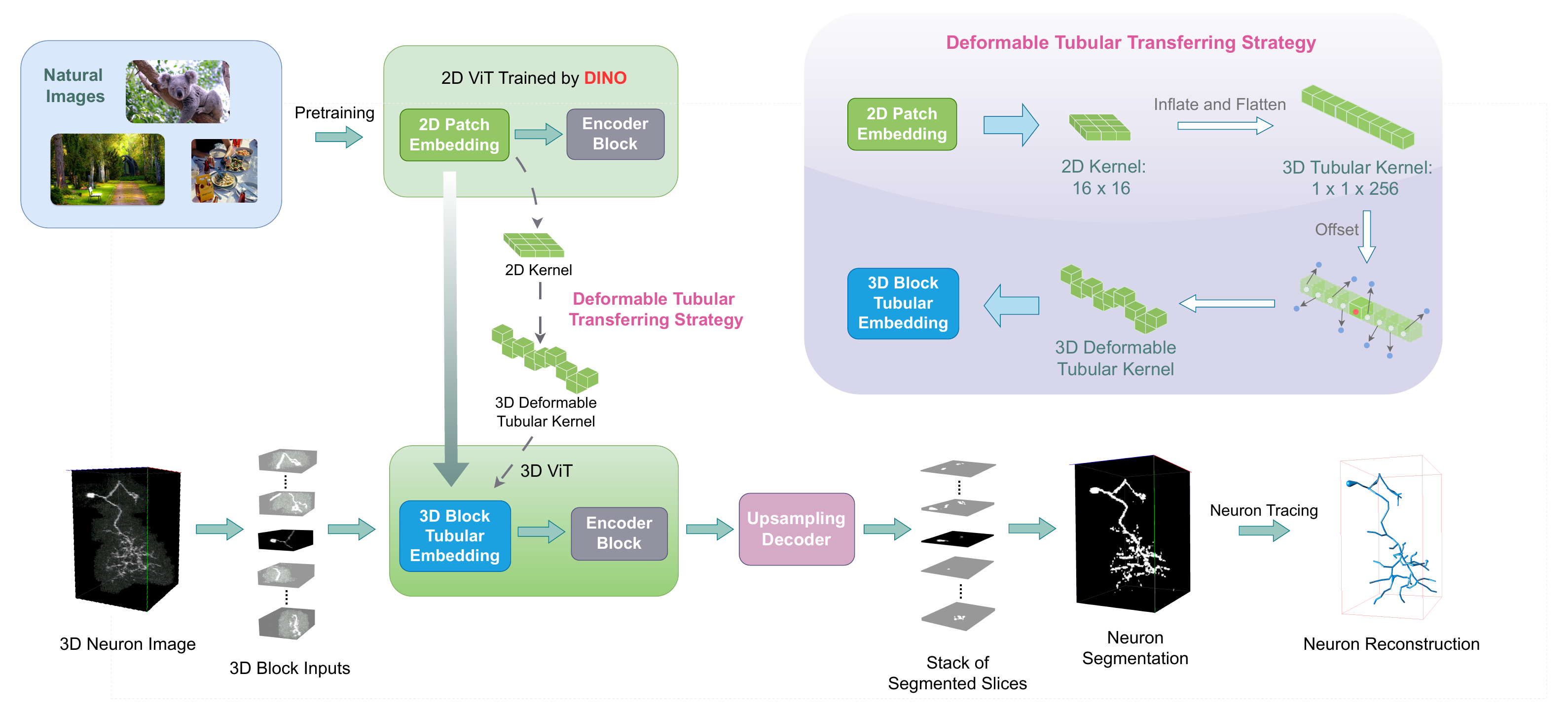}
\caption{The overview of our proposed framework, \textbf{\textit{DINeuro}}. The 3D neuron image is initially partitioned into multiple 3D blocks, which are forwarded to a 3D ViT for block-to-slice segmentation. The 3D ViT is initialized through our deformable tubular transferring strategy, which dynamically aligns the pre-trained 2D kernel into a 3D deformable tubular kernel to better capture the tree-shape features of a 3D neuron image. The segmented slices are then stacked to produce the final segmentation, prior to applying a tracing method to reconstruct the overall neuron structure.} \label{fig1}
\end{figure}

In this work, we design an innovative framework that leverages the prior knowledge from \underline{\textbf{\textit{DIN}}}O~\cite{caron2021emerging} to facilitate n\underline{\textbf{\textit{euro}}} morphological learning, namely \underline{\textbf{\textit{DINeuro}}}. This proposed method captures the inherent tree-shape properties of neuron structures. It is achieved by transferring the knowledge from 2D natural images into our 3D Vision Transformer to capture the topological and geometric features of 3D neuron images. To bridge the gap between 2D natural images and 3D neuronal topology, we design a deformable tubular transferring strategy to dynamically distill the 2D natural knowledge to the varied tubular characteristics of neuron morphology. In summary, our contributions are as follows: 

\begin{itemize}
    \item We propose to leverage 2D natural image knowledge for distinct neuronal topological learning.  
    
    \item We develop a novel framework, \textbf{\textit{DINeuro}}, to build the knowledge connection between the 2D natural knowledge and intrinsic neuron morphological features through a deformable tubular transferring strategy. This strategy distills the 2D consensus knowledge from DINO~\cite{caron2021emerging} and adapts it to the tubular characteristics of neurons within a multi-view 3D deformable tubular network. 
    
    % \item We build the knowledge connection between the 2D natural knowledge and intrinsic neuron morphological features, enabling more effective adaptation of 2D natural knowledge to distinct 3D neuron structures. 
    
    \item Our framework significantly achieves segmentation performance improvement of 4.53\% in mean Dice and 3.56\% in mean 95\% Hausdorff distance, learning the intricate tree-like structure of 3D neuron images more efficiently and effectively.
\end{itemize}

\section{Methodology}
\subsection{Network Architecture and Weight Inflation Transfer Method from 2D to 3D}
The overview of our network architecture is illustrated in Figure 1. Our approach builds upon transferring knowledge obtained from DINO~\cite{caron2021emerging}, a self-supervised 2D Vision Transformer (ViT) model trained on extensive 2D natural image datasets, into a 3D deformable tubular network. To address the dimensional disparity between the 2D knowledge learned from DINO and our 3D segmentation model~\cite{cheng2024boosting}, we adopt the weight transfer technique introduced in work~\cite{cheng2024boosting} at the first stage. This method inflates the 2D weights from DINO into a 3D space through average and center strategy, which spatially extends the receptive field of the 2D weights. The inflated weights are then used to initialize our 3D network, ensuring the features learned from the 2D natural image domain can be effectively applied to 3D volumetric neuron data. 

\subsection{Deformable Tubular Weight Transferring Strategy}
Although the transferring strategy in Section 2.1 effectively brings the knowledge from 2D natural images into the 3D neuron image domain, the transferred 2D kernels are limited by their original design that focuses on the extraction of local features, thus neglecting the varied tree-shape structural properties of neurons. To overcome this limitation, inspired by~\cite{qi2023dynamic}, we propose a novel framework, \textbf{\textit{DINeuro}}, with a deformable tubular weight transferring strategy. As shown in Figure 1, this strategy transfers the 2D natural knowledge obtained from DINO. The transferred prior knowledge is then adapted to enhance the morphological learning of the inherent tree-like features of volumetric neurons. This adaptation is achieved by deforming the transferred 2D convolutional kernels into tubular forms. 

\subsubsection{Transferring 2D Weights into 3D Tubular Kernel}
Our proposed 3D deformable tubular weight transferring strategy enhances the adaptability of the transferred 2D convolutional kernel through reshaping the perceptual field into a tubular shape to effectively capture the complex geometric and pipe-like features of neurons. In the embedding layer of DINO, the 2D kernel is with the size of $16\times16$. To adapt this kernel to the tubular feature of 3D neurons, as illustrated in Figure 2, we flatten the 2D kernel to a 3D tubular kernel with the shape of $1\times1\times256$.  

\begin{figure}
\includegraphics[width=11.90cm]{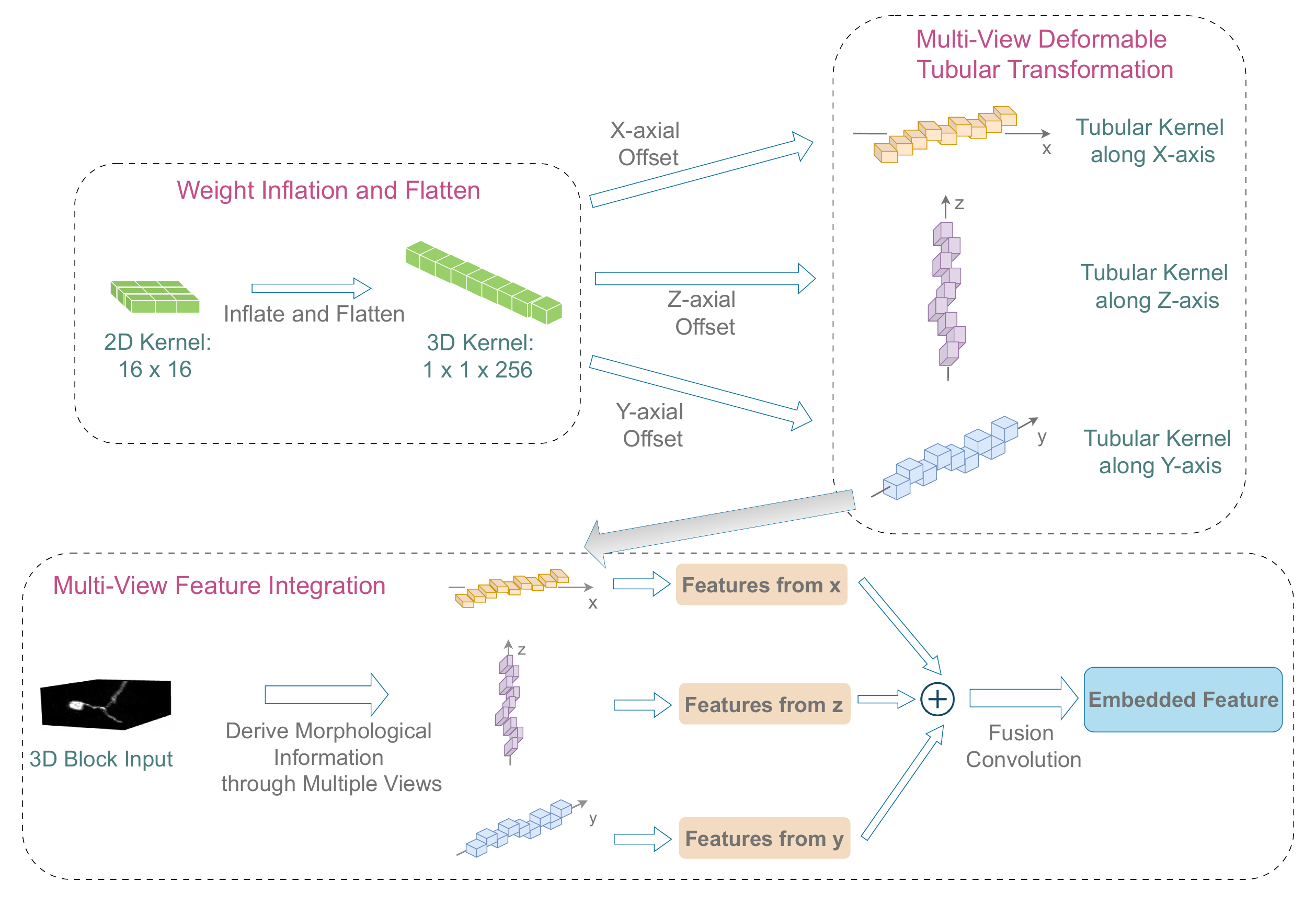}
\caption{Illustration of the fusion of extracted multi-view features along the z, y, and x directions. The 2D convolutional kernel is adapted and shifted to 3D deformable tubular kernels in three directions, which are used to capture the multi-view geometric information of 3D neurons.} \label{fig2}
\end{figure}

\subsubsection{Multi-View Tubular Adaptation and Iterative Tubular Offset}
Neuronal structures are usually highly anisotropic with varying diameters and angles. To effectively extract diverse geometric and topological information from different perspectives of volumetric neuron images, as presented in Figure 2, we propose a multi-view learning strategy. This strategy flattens the transferred 2D convolutional kernel into three distinct 3D tubular kernels along the z-axis, y-axis, and x-axis respectively. Deformation offsets are then applied to shift the tubular kernels and align them with the tubular morphology of neurons, enabling better capture of neuronal branching structures. Taking the kernel transferred along the x-axis as an example, the transferred 2D convolutional kernel from DINO is inflated and flattened in the direction of the x-axis to form a 3D kernel in a tubular shape. The length of this tubular kernel is $16\times16$, with the x-coordinates of the grids being continuous and fixed. The x-coordinates of the grids in this kernel can be represented as: $x_{0+h}$, where $x_0$ is the central grid and $h = \{-127, .., 128\}$, denoting the horizontal distance of each grid from the central grid. The y-coordinate and z-coordinate for each grid in this tubular convolution kernel are generated by iterative offsets. Through iteratively producing an offset based on the offset of the preceding grid point within the kernel and constraining the step of each offset, we ensure a coherent and continuous transformation of the kernel into a tubular shape and effectively minimize the risk of significant deviation. A grid position $(x_i, y_i, z_i)$ is dependent on the previous grid $(x_{i-1}, y_{i-1}, z_{i-1})$ and can be described by: $y_i = y_{i-1} + \Delta y_{i}, \quad z_i = z_{i-1} + \Delta z_{i}$, where $\Delta y_{i}$ and $\Delta z_{i}$ denote the offsets along the y-axis and z-axis for $y_i$ and $z_i$ respectively, ranging from $[-1, 1]$. Therefore, the offsets for the y-coordinate and z-coordinate are iterative and accumulated, assuring the transferred 2D kernel is aligned with a continuous tubular neuron structure. 

\section{Experiments and Results}
\subsection{Dataset and Experiment Settings} 

Our experiments are conducted on the publicly available Janelia dataset of the BigNeuron project~\cite{peng2015bigneuron,manubens2023bigneuron}. The dataset is divided into training, validation, and testing sets, comprising 38, 3, and 4 images respectively. Due to computational resource constraints, the volumetric neuron images are divided into smaller 3D blocks with dimensions of $100\times100\times5$. Only the blocks containing a significant number of neuronal voxels (determined by a predefined foreground ratio) are selected for training. Consistent with previous work~\cite{wang2019segmenting}, we generate the ground truth for neuron segmentation by utilizing a scale-space distance transformation technique.

\subsection{Results}

To quantitatively evaluate the segmentation performance of our proposed framework, we employ two widely used metrics in the field of image segmentation: mean Dice and mean 95\% Hausdorff distance (Hd95). The SmartTracing method~\cite{chen2015smarttracing} is applied as the tracing algorithm to reconstruct neuron morphology from segmented results. Three spatial distance-driven metrics—Entire Structure Average (ESA), Different Structure Average (DSA), and Percentage of Different Structure (PDS)—are used to measure the geometric discrepancy between the reconstructed neuron and the manually annotated ground truth. The visualizations of segmentation results and reconstruction results for a sample are displayed in Figure 3. 

\begin{figure}
\includegraphics[width=11.75cm]{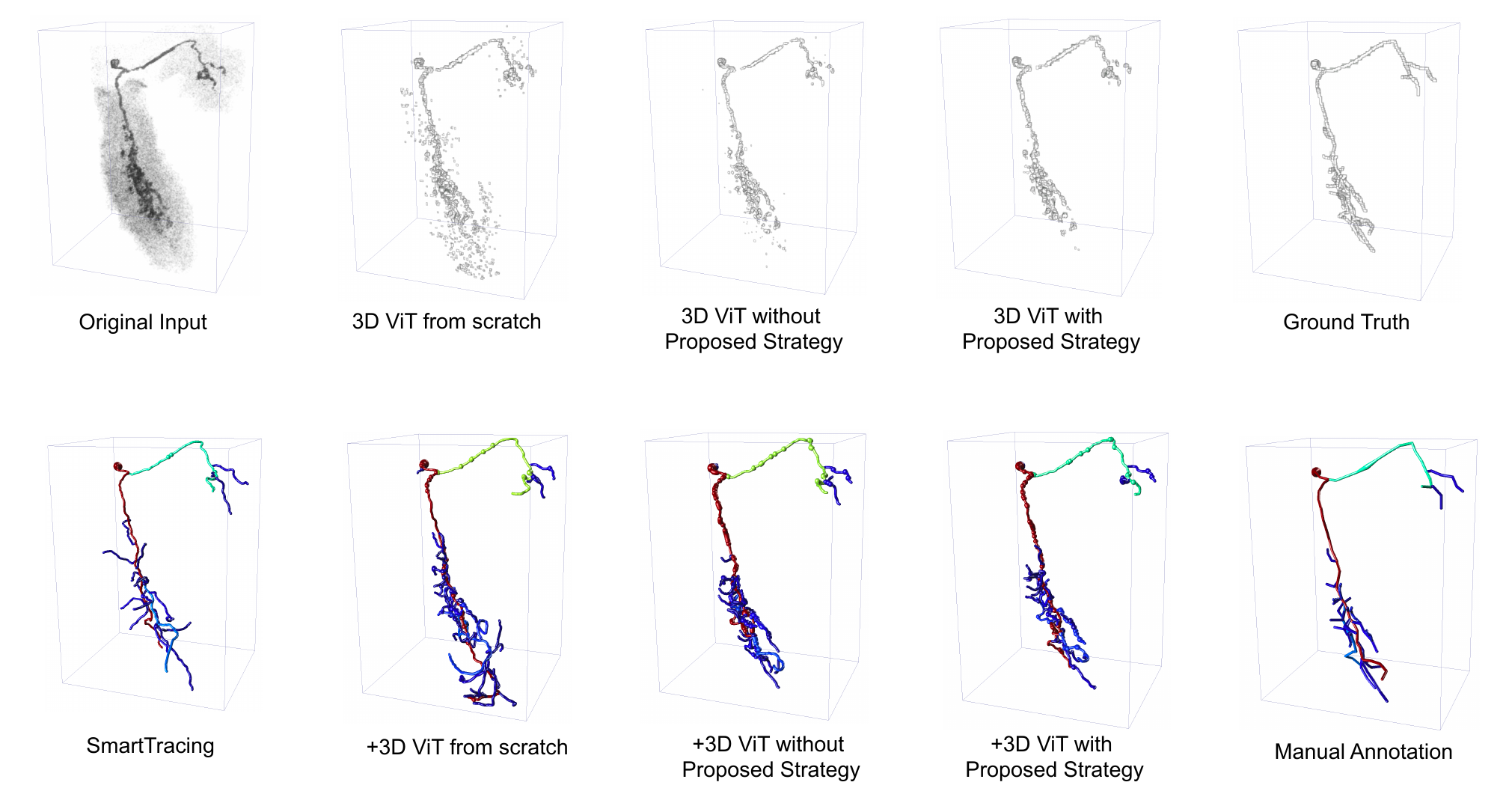}
\caption{Illustration of the segmentation results (top) and corresponding neuron reconstruction (bottom) through SmartTracing algorithm for a sample 3D neuron image. The '+' symbol indicates the application of SmartTracing to the segmented outputs from the model.} \label{fig3}
\end{figure}

\renewcommand{\arraystretch}{1.3}
\begin{table}[H]
\footnotesize  % Adjust font size
\captionsetup{font=small}
\caption{Segmentation performance of 3D ViT with and without deformable tubular transferring strategy.} 
\vspace{10pt}  % Adjust this value as needed
\centering
\begin{tabular}{>{\centering\arraybackslash}m{1.5cm} >{\centering\arraybackslash}m{2.5cm} >{\centering\arraybackslash}m{2.5cm}| >{\centering\arraybackslash}m{1.5cm} >{\centering\arraybackslash}m{1.5cm}}  % Use m for vertical centering
\hline
Model & Deformable Tubular Transferring & 2D to 3D Weight Transfer Method & Mean Dice$\uparrow$ & Mean Hd95$\downarrow$ \\
\hline
\hline
\multirow{5}{*}{3D ViT} & \ding{55} & /       & 0.4174 & 37.943 \\
                        & \ding{55} & Average & 0.4942 & 2.863 \\
                        & \ding{55} & Center  & \textbf{0.5045} & \textbf{2.810} \\
                        & \ding{51} & Average & 0.5455 & 2.840 \\
                        & \ding{51} & Center  & \textbf{0.5498} & \textbf{2.710} \\
% \hline
% \multirow{2}{*}{3D ViT} & \ding{51} & Average & 0.5455 & 2.84 \\
%                         & \ding{51} & Center  & \textbf{0.5498} & \textbf{2.71} \\
\hline
\end{tabular}
\label{tab:example}
\end{table}

\subsubsection{Neuron Segmentation Results}

The quantitative results of the neuron segmentation task are presented in Table 1, showcasing the superior performance of our proposed framework. As shown in Table 1, through incorporating tubular characteristics of neuron morphology into the model, the segmentation performance improves significantly. The mean Dice score increases from 0.5045 to 0.5498, and the mean Hd95 improves from 2.810 to 2.710, compared to the model without applying proposed deformable tubular transferring strategy. Moreover, models employing center transferring strategy demonstrate further improvement. 

\renewcommand{\arraystretch}{1.3}
\begin{table}[H]
\footnotesize  % Adjust font size
\captionsetup{font=small}
\caption{Reconstruction performance of 3D ViT with and without deformable tubular transferring strategy.}
\vspace{10pt}  % Adjust this value as needed
\centering
\begin{tabular}{>{\centering\arraybackslash}m{1.5cm} >{\centering\arraybackslash}m{2.5cm}| >{\centering\arraybackslash}m{1.5cm}| >{\centering\arraybackslash}m{1.5cm}| >{\centering\arraybackslash}m{1.5cm}}  % Narrower columns
\hline
Model & Deformable Tubular Transferring & ESA$\downarrow$ & DSA$\downarrow$ & PDS$\downarrow$ \\
\hline
% \hline
\multirow{2}{*}{3D ViT} & \ding{55} & 2.23 & 4.78 & 0.27 \\
& \ding{51} & \textbf{2.07} & \textbf{4.10} & \textbf{0.26} \\
                        
% \hline
% \multirow{1}{*}{3D ViT} & \ding{51} & 2.0732 & 4.1001 & 0.2565 \\
                        
\hline
\end{tabular}
\label{tab:example}
\end{table}

\subsubsection{Neuron Reconstruction Results} 
To further validate the performance of our proposed framework in neuron reconstruction, we employ the SmartTracing algorithm~\cite{chen2015smarttracing} to trace the neuronal morphology from the segmented outputs. The tracing performance of our framework is presented in Table 2. Notably, our proposed framework achieves the best quantitative reconstructed results across all metrics. A visualization comparison of reconstruction results is presented in Figure 3. It is notable to find that the reconstructed neuron generated by our deformable tubular transferring strategy exhibits fewer artifacts and noises, producing a much cleaner and more accurate representation that aligns more closely with the ground-truth annotation. 

\section{Conclusion}
In this work, we propose a novel framework, \textbf{\textit{DINeuro}}, that significantly enhances segmentation performance through facilitating 3D neuronal morphological learning with 2D natural knowledge. Our work demonstrates the effectiveness of incorporating the geometric topological information of neuron structure for precise neuron reconstruction and provides a solution to establish a knowledge connection between 2D natural knowledge and neuron morphological properties. Through leveraging the prior knowledge from DINO and adapting it to a 3D deformable tubular network, our proposed framework adaptively extracts the tree-shape features of varied neuron morphology. The results from our experiments indicate that our proposed strategy significantly improves segmentation performance, with a 4.53\% increase in mean Dice score and a 3.56\% enhancement in mean Hd95.

\bibliographystyle{splncs04}
\bibliography{strings}




% LNCS Homepage, \url{http://www.springer.com/lncs}, last accessed 2023/10/25
% \end{thebibliography}
\end{document}